\documentclass[11pt]{article}

\usepackage{fullpage, hyperref,bbm,amssymb,amsfonts,amsthm}
\usepackage[noend]{algorithmic}
\usepackage{algorithm}
\usepackage{algorithm}
\usepackage{setspace}
\usepackage{color}

\newcommand{\C}{\mathbb{C}}

\newtheorem{theorem}{Theorem}

\newtheorem{cor}{Corollary}

\begin{document}
\title{Sensitivity of Quantum Walks with Perturbation} \author{
Chen-Fu Chiang\thanks{School of Electrical Engineering
and Computer Science,
University of Central Florida, Orlando, FL~32816, USA. Email:
\texttt{cchiang@eecs.ucf.edu}}
\quad
}		
		
\maketitle

\begin{abstract}
Quantum computers are susceptible to noises from the outside world. We
investigate the effect of perturbation on the
hitting time of a quantum walk and the stationary distribution prepared by a
quantum walk based algorithm. The perturbation comes from quantizing
a transition matrix Q with perturbation E (errors). We bound the perturbed 
quantum walk hitting time from above by applying Szegedy's work and the
perturbation bounds with Weyl's perturbation theorem on classical matrix. Based
on an efficient quantum sample preparation approach invented in {\em speed-up
via quantum sampling} and the perturbation bounds for stationary distribution
for classical matrix, we find an upper bound for the total variation
distance between the prepared quantum sample and the true quantum sample.
\end{abstract}

\section{Introduction}
Markov chains and random walks have been useful tools in classical
computation. One can use random walks to obtain the final stationary 
distribution of a Markov chain to sample from. In such an application the 
time the Markov chain takes to converge, i.e., {\em convergence time}, is
of interest because shorter convergence time means lower cost in
generating a sample. Sampling from stationary distributions of Markov chains combined with simulated 
annealing is the core of many clever classical approximation algorithms. 
For instance, approximating the volume of convex bodies \cite{LV:06}, approximating the 
permanent of a non-negative matrix \cite{JSV:04}, and the partition function of
statistical physics models such as the Ising model \cite{JS:93} and the Potts model \cite{BSVV:08}. 
In addition, one can also use the random walks to search for the {\em{marked}}
state in the Markov chain, in which the {\em hitting time} is of interest. It
is because hitting time indicates the time it requires to find the marked state. \\

The Markov Chain Monte Carlo (MCMC) method is a centerpiece of many efficient
classical algorithms. It allows us to approximately sample from a
particular distribution $\pi$ over a large state space $\Omega$.  Perturbations
of classical Markov chains are widely studied with respect to hitting time
and stationary distribution. The variation of a stationary distributuion
caused by perturbation would deteriorate the accuracy of our sampling. With the
above facts, we are interested to know what effect perturbation has on
currently existing quantum walk based algorithms. \\

The note is organized as follows. 
In section \ref{ClassicalGap} we present the deviation effect of perturbation
on spectral gap of Markov chain and we apply it to the hitting time of a quantum walk in section
\ref{hittingtimeDeviation}. In section \ref{ClassicalSampPerturb} we state the result of total variation distance of
classical stationary distributions. By using an efficient algorithm \cite{WA:08}
for preparing quantum samples, in section \ref{distributionDeviation} we obtain
the total variation distance between the prepared quantum sample and the true
quantum sample. Finally in section \ref{conclusion}, we make our conclusion and suggest future works.

\section{The Perturbation}\label{Deviation}
Given a stochastic symmetric matrix $P \in \C^{n \times n}$, we can quantize the
Markov chain \cite{Szegedy:04} and we showed that the implementation of
one step of quantum walk \cite{CNW:10} can be achieved efficiently. However,
the above settings always are under the assumption of perfect
scenarios. In real life there are many sources of errors that would perturb
the process. Noises might be propagated along with the input source or
they might be introduced during the process. Here we solely look at the noises
that are introduced at the beginning of the process. \\

The noises can be introduced due to the precision limitation and the noisy
environment. For instance, not all numbers have a perfect binary
representation and the approximated numbers would cause perturbation. Suppose 
our input decoding mechanism can always take the input matrix and represent it
in a symmetric transition matrix $Q$ where $Q$ can be perfectly represented by
the limited precision and it is the matrix closes to the original matrix $P$
that the system can prepare. \\

Let $E$ be the noise that is introduced because of system's precision
limitation and the environment, we can express the transistion matrix as 
\begin{equation}
   Q =  P + E. 
\end{equation} 
\noindent

\subsection{Classical Spectral Gap Perturbation}\label{ClassicalGap}
Classically many researches \cite{IN:09, CM:01, GL:96, Parlett:98,
BF:60,EI:99} are focused on the spectral gaps and stationary distributions 
of the matrice with perturbation. In a recent work by Ipsen and
Nadler \cite{IN:09} , they refined the perturbation bounds for eigenvalues of
Hermitians. Throughout the rest of the note, $\| \cdot \|$ always denotes
$l_2$ norm, unless otherwise specified. Based on their result, we summarized the following:


\begin{cor}\label{WeylPerturb}
Suppose $P$ and $Q$ $\in \C^{n \times n}$ are Hermitian symmetric transition
matrices with respective eigenvalues
\begin{equation}
\lambda_n(P) \leq \ldots \leq \lambda_1(P) = 1, \quad \quad \lambda_n(Q) \leq
\ldots \leq \lambda_1(Q) = 1,   
\end{equation}
and $Q = P + E$, then 
\begin{equation}\label{eqn:WeylPerturb}
\max_{1\leq i \leq n} |\lambda_i(P) - \lambda_i(Q)| \leq \|E \|. 
\end{equation}

\noindent
Furthermore, the spectral gap $\delta$ of $P$ and the spectral gap $\Delta$ of
$Q$ have the following relationship
\begin{equation}
\delta - \|E\| \leq \Delta \leq \delta + \|E\|.
\end{equation}
\end{cor}
\proof
Eq.~(\ref{eqn:WeylPerturb}) is a direct result from the Weyl's
Perturbation Theorem. The {\em Weyl's Perturbation Theorem} bounds the worst-case absolute error between the $i$th exact and
the $i$th perturbed eigenvalues of Hermitian matrices in terms of the
$l_2$ norm \cite{GL:96, Parlett:98}.
And since
\begin{equation} 
	1 - \lambda_2(P) = \delta, \quad\quad 1 - \lambda_2(Q) = \Delta,\nonumber
\end{equation}
by eq.~(\ref{eqn:WeylPerturb}) we have $|\delta -
\Delta| \leq \|E\|$. Therefore, in general we can bound the perturbed
spectral gap $\Delta$ as
\begin{equation}
\delta - \|E\| \leq \Delta \leq \delta + \|E\|. \nonumber 
\end{equation} \qed

Generally speaking, the global norm of $E$ might be very large when the
dimensions $n >> 1$ \cite{Johnstone:01}. However, in our case because $E$ is the
difference between two very close stochastic symmetric matrices, its
global norm would never become large.

\subsection{Quantum Hitting Time Perturbation}\label{hittingtimeDeviation}
Given two Hermitian stochastic matrices, $P$ and $Q$, we 
explore the difference between walk operators, $W(P)$ and $W(Q)$, with respect
to their hitting time. Denote the set of marked elements as
$|M|$. Based on the result from {\em Corollary \ref{WeylPerturb}}, we have the
following:

\begin{cor}
Given two symmetric reversible ergodic transition matrices $P$ and
$Q$ $\in \C^{n \times n}$, where $Q = P + E$, let $W(P)$ and $W(Q)$ be
quantum walks based on $P$ and $Q$, respectively. Let $M$ be the set of marked
elements in the state space. Denote $HT(P)$  the hitting time of walk $W(P)$
and $HT(Q)$ the hitting time of walk $W(Q)$. Suppose $|M| = \epsilon N$. If the
second largest eigenvalues of $P$ and $Q$ are at most $1 - \delta$ and $1-\Delta$, respectively, 
then in general

\begin{equation}
HT(P) = O\Big(\sqrt{\frac{1}{\delta \epsilon}}\Big), \quad \quad HT(Q) =
O\Big(\sqrt{\frac{1}{(\delta-\|E\|)  \epsilon}}\Big)
\end{equation}
where $ \delta -\|E\| \leq \Delta \leq \delta + \|E\|$. 


\proof Suppose the Markov chain $P$, $Q$ and matrix $E$ are in the following
block structure
\begin{equation}
P = \left( \begin{array}{cc}
P_1 & P_2 \\
P_3 & P_4 \end{array} \right) , \quad
Q = \left( \begin{array}{cc}
Q_1 & Q_2 \\
Q_3 & Q_4 \end{array} \right), \quad
E = \left( \begin{array}{cc}
E_1 & E_2 \\
E_3 & E_4 \end{array} \right)
\end{equation}
where we order the elements such that the marked ones come last, i.e., $P_4$,
$Q_4$ and $E_4 \in \C_{|M| \times |M|}$. The corresponding modified Markov
chains \cite{Szegedy:04} would be
\begin{equation}
\tilde{Q} = \left( \begin{array}{cc}
Q_1 & 0 \\
Q_3 & I \end{array} \right) = \left( \begin{array}{cc}
P_1 + E_1 & 0 \\
P_3 + E_3 & I \end{array} \right).
\end{equation} 

By \cite{Szegedy:04}, we have $HT(P) = O(\sqrt{\frac{1}{1-\|P_1\|}})$
and $HT(Q) = O(\sqrt{\frac{1}{1-\|Q_1\|}})$. Since we know
\begin{equation} \|P_1\| \leq 1 - \frac{\delta \epsilon}{2} 
\quad \mbox{and} \quad \|Q_1\| \leq 1 - \frac{\Delta \epsilon}{2} 
\end{equation}
by \cite{Szegedy:04} and by
Cauchy's interlacing theorem we have  $\|E\| \geq \|E_1\|$
\cite[Cor.III.1.5]{Bhatia:97}, we then obtain 
\begin{equation}\|Q_1\| \leq \min\left\{\|P_1\| + \|E\|, 1-\frac{(\delta
-\|E\|)\epsilon}{2}\right\} 
\end{equation}
as $ \delta -\|E\| \leq \Delta \leq \delta + \|E\|$.
Therefore, the hitting times for $P$ and $Q$ are derived.
\qed
\end{cor}

\subsection{Classical Sample Perturbation}\label{ClassicalSampPerturb} 
In this section we adapt the results from the work
\cite{CM:01} to bound the stationary distriubtion $\pi(Q)$ of a perturbed matrix $Q$ with respect to the
perturbation $E$ and the true stationary distribution $\pi(P)$, i.e.,
\begin{equation}\label{columnstochastic}
Q \cdot \pi(Q) = \pi (Q), \quad P \cdot \pi(P) = \pi(P). 
\end{equation} 
Let $\Omega$ be the state space and $\Omega' = \Omega \cup \{0\}$. The {\em
total variation distance} between two probability distributions over $\Omega$
is defined as
\begin{equation}
D(\pi(P), \pi(Q)) = \frac{1}{2}\sum_{x \in \Omega}\|\pi(P)_x - \pi(Q)_x\|_1 =
\max_{S \subseteq \Omega'}|\pi(P)_S - \pi(Q)_S|.
\end{equation}

\noindent
Here $\pi(P)$ denotes the stationary distribution of matrix $P$, $\pi(P)_x$
is the $x$th element of $\pi(P)$ and $\pi(P)_S$ denotes the sum of
$\pi(P)_x$ where $x \in S$, i.e., $\sum_{x \in S}\pi(P)_x = \pi(P)_S$.  \\

In \cite{CM:01} it is assumed that
the transition matrix is {\em row-wise stochastic}. Our matrix is column-wise stochastic (see
eqn.(\ref{columnstochastic})) but since it is symmetric, it is also
row-stochastic. By chooseing {\em condition number} $\kappa_5$ in \cite{CM:01},
the {\em ergodicity coefficient}, using the $l_p$ norm, is defined as
\begin{equation}
\tau_p(P) = \sup_{\|v\|_p = 1,\\ v^T e = 0}\|v^T P\|_p
\end{equation} 

\noindent
where $e$ is a column vector of all ones. Since $P$ is a stochastic matrix,
the ergodic coefficient satisfies $0 \leq \tau_1(P) \leq 1$. In case of
$\tau_1(P) < 1$, we have a perturbation bound  in terms of the ergodic
coefficient of $P$:

\begin{equation}{\label{ClassicalPerturbedDistBound}}
D(\pi(P), \pi(Q)) = \frac{1}{2}\|\pi(P) - \pi(Q)\|_1 \leq
\frac{1}{2(1-(\tau_1{P}))}\|E\|_{\infty}.
\end{equation}

\subsection{Quantum Sample Perturbation}\label{distributionDeviation}
While there are several methods that make use of Szegedy's quantum walk
operators to prepare quantum samples \cite{WA:08, SBB:07, SBBK:08}, we choose
\cite{WA:08} as the main approach to analyze as it leads to an overall speed-up
in the general case. The other approaches \cite{SBB:07, SBBK:08} take
advantage of the quantum Zeno effect but the problem is that the
quantum Zeno effect would result in an exponential slow-down in the general case. \\

The work by Wocjan and Abeyesinghe \cite{WA:08} showed an approach to
prepare the coherent stationary distribution of a Markov Chain via modified quantum walk
and Grover's $\frac{\pi}{3}$-amplitude amplication techniques. The theorem
listed below is the main theorem in {\em Speed-up via Quantum Sampling}. We refer the
interested readers to \cite{WA:08} for details on this algorithm for the
construction techniques and the computational complexity.

\begin{theorem}[Speed-up via quantum sampling
\cite{WA:08}]{\label{SpeedupQSample}} {\label{QSampleConstruct}}Let $Q_0, Q_1,
\ldots., Q_r = Q$ be a sequence of classical Markov chains with stationary
distributions $\pi_0, \pi_1, \ldots, \pi_r$ and spectral gap $\delta_0, \ldots,
\delta_r$. Assume that the stationary distributuions of adjacent Markov chains
are close to each other in the sense that $|\langle \pi_i |\pi_{i+1}\rangle|^2 \geq c$ 
where c is some constant, for $i=0,\ldots,r-1.$. Then for any $\eta > 0$, there is an 
efficient quantum sampling algorithm, making it possible to sample according to a 
probability distributuion $\tilde{\pi}_r$ that is close to $\pi_r$ with respect to the
total variation distance, i.e., $D(\tilde{\pi}_r, \pi_r) \leq \eta$.
\end{theorem}

Based on the theorem above, we can immediately conclude the following corollary: 
\begin{cor}
When the coherent quantum sample based on the perturbed
Markov chain is prepared by using techniques of \cite {WA:08}
with precision $\eta$, the total variation distance between the
prepared quantum sample $\tilde{\pi}(Q)$ and the true quantum sample $\pi(P)$
is less than $\eta + \frac{1}{2(1-(\tau_1{P}))}\|E\|_{\infty}$.
\end{cor}
\proof
By {\em Theorem \ref{SpeedupQSample}} we can
efficiently construct a quantum sample $\tilde{\pi}{(Q)}_r$ that is $\eta$
close to $\pi(Q)$. Then by triangle inequality we obtain
\begin{equation}
D(\pi(P), \tilde{\pi}(Q)) \leq D(\pi(P), \pi(Q)) + D(\pi(Q) + \tilde{\pi}(Q))
\leq \frac{1}{2(1-(\tau_1{P}))}\|E\|_{\infty} + \eta .  
\end{equation}
\qed


\section{Conclusion and Discussion}\label{conclusion}
We apply the existing classical results from matrix perturbation 
to quantum walk based algorithms. With the noise introduced at the input, as
quantum system is susceptible to the outside world and some other precision
limitation problem, we bounded the quantum hitting time with perturbation from
the above that the perturbed quantum walk hitting time is  
\begin{equation}
HT(Q) =
O\Big(\sqrt{\frac{1}{(\delta-\|E\|)  \epsilon}}\Big). \nonumber
\end{equation}

In the meanwhile, we also showed that how the quantum sample
prepared by using the approach in \cite{WA:08} would fluctuate from the true
quantum sample when perturbation is present. The analysis is based on the
assumption that we have a series of Markov chains $Q1, \ldots, Q_r = Q$. Hence,
we have 

\begin{equation}
D(\pi(P), \tilde{\pi}(Q)) \leq  \frac{1}{2(1-(\tau_1{P}))}\|E\|_{\infty} +
\eta . \nonumber
\end{equation}

Intuitively from the analysis we can see that the total variation distance for
$D(\pi(\tilde{Q}), \pi(Q))$ is simply {\em additive} and $D(\pi(P), \pi(Q))$
cannot be eliminated. However, if the matrix $P = Q_i$ is inside the sequence
$Q_1, \ldots, Q_r$ where $1 < i < r$, can we invent a procedure to detect to
avoid such overshoot? Future study is to find the relation between
quantum mixing time, the time it takes to get $\eta$-close to the true
stationary distribution, and the quantum hitting time as it was studied so classically.
Furthermore, another possible analysis approach would be
assuming that we have a series of Markov chains $P_1, \ldots, P_r = P$ (without
the noise). We can adapt the analysis in \cite{WA:08} to study how the noise would
affect (i) accuracy when blindly preparing the quantum sample without
acknowledging the existence of noise or (ii) complexity when the noise is
acknowledged and desired accuracy must be achieved. \\

\section{Acknowledgments}
		C.~C. gratefully acknowledges the support of NSF grants
		CCF-0726771 and CCF-0746600.


\begin{thebibliography}{10}
	 \bibitem{LV:06}
		L.~Lov\'{a}sz and S.~Vempala, {\em Simulated Annealing in Convex Bodies and an $O^*(n^4)$ Volume Algorithm},
		Journal of Computer and System Sciences, vol.~72, issue 2, pp. 392--417, 2006.

	\bibitem{JSV:04}
		M.~Jerrum, A.~Sinclair, and E.~Vigoda, {\em A Polynomial-Time
		Approximation Algorithm for the Permanent of a Matrix Non-Negative
		Entries}, Journal of the ACM, vol.~51, issue 4, pp.~671--697, 2004.

	\bibitem{JS:93}
		M.~Jerrum and A.~Sinclair, {\em Polynomial-Time Approximation
		Algorithms for the Ising Model}, SIAM Journal on Computing, vol.~22,
		pp.~1087--1116, 1993.

	\bibitem{BSVV:08}
		I.~Bez\'{a}kov\'{a}, D.~\v{S}tefankovi\v{c}, V.~Vazirani and E.~Vigoda,
		{\em Accelerating Simulated Annealing for the Permanent and
		Combinatorial Counting Problems}, SIAM Journal on Computing, vol.~37,
		No. 5, pp.~1429--1454, 2008.
	
	\bibitem{WA:08}
		P.~Wocjan and A.~Abeyesinghe, {\em Speed-up via Quantum Sampling},
		Phys. Rev. A, vol.~78, pp.~042336, 2008.
	
	 \bibitem{Szegedy:04}
		M.~Szegedy, {\em Quantum Speed-up of Markov Chain Based Algorithms},
		Proc. of 45th Annual IEEE Symposium on Foundations of Computer
		Science, pp.~32--41, 2004.
		
	\bibitem{CNW:10}
		C.-F.~Chiang, D.~Nagaj, P.~Wocjan, {\em Efficient Circuits for the Quantum
		Walks}, QIC vol.~10 no.~5\&6 pp.~0420--0434, 2010.
			
	\bibitem{IN:09}
        I.~Ipsen and B.~Nadler, {\em Refined Perturbation Bounds for Eigenvalues
        of Hermitian and Non-Hermitian Matrices}, SIAM J. Matrix Anal. Appl.,
        vol.~31, no.~1, pp.~40--53, 2009.
		
    \bibitem{CM:01}
       G.~Cho and C.~Meyer, {\em Comparison of Perturbation Bounds for the
       Stationary Distribution of a Markov Chain}, vol.~335, issue 1-3, pp.137 -
       150, Linear Algebra and Its Applications, 2001.
		
	\bibitem{GL:96}
       G.~Golub and C.~Loan, {\em Matrix Computations}, 3rd ed., The Johns
       Hopkins University Press, 1996. 
       
    \bibitem{Parlett:98}
       B.~Parlett, {\em The Symmetric Eigenvlaue Problems}, SIAM, Philadelphia,
       1998. 
       
    \bibitem{BF:60}
        F.~Bauer and C.~Fike, {\em Norms and Exclusion Theorems}, Numer.
        Math.,vol.~2, pp.~137 - 141, 1960. 	
        
      \bibitem{EI:99}
        S.~Eisenstat and I.~Ipsen, {\em Three Absolute Perturbation Bounds for
        Matrix Eigenvalues Imply Relative Bounds}, SIAM Journal on Matrix
        Analysis and Applications, vol.~20 ,  issue 1, pp.~149 - 158, 1999.
		
	 \bibitem{Johnstone:01}
       I.~Johnstone, {\em On the distribution of the largest eigenvalue in
       principal components analysis}, vol.~29, no.~2, pp.~295 - 327, Annals of
       Statistics, 2001.
    
    \bibitem{Bhatia:97}
        R. Bhatia, {\em Matrix Analysis}, Springer Verlag, New York, 1997.     
        
    \bibitem{SBB:07}
		R.~Somma, S.~Boixo, and H.~Barnum, {\em Quantum Simulated Annealing},
		arXiv:abs/0712.2008
		
	\bibitem{SBBK:08}
		R.~Somma, S.~Boixo, H.~Barnum, E.~Knill, {\em Quantum Simulations of
		Classical Annealing Processes}, Phys. Rev. Lett. vol.~101, pp.~130504, 2008.
       

   
\end{thebibliography}
	 \end{document}